\documentclass{WileyMSP-template}
\usepackage{color,soul}
\usepackage{textcomp}
\begin{document}

\pagestyle{fancy}

\title{Unravelling and circumventing failure mechanisms in chalcogenide optical phase change materials}

\maketitle


\author{Cosmin Constantin Popescu},
\author{Kiumars Aryana},
\author{Brian Mills},
\author{Tae Woo Lee},
\author{Louis Martin-Monier},
\author{Luigi Ranno},
\author{Jia Xu Brian Sia},
\author{Khoi Phuong Dao},
\author{Hyung-Bin Bae},
\author{Vladimir Liberman},
\author{Steven Vitale},
\author{Myungkoo Kang},
\author{Kathleen A. Richardson},
\author{Carlos A. R\'ios Ocampo},
\author{Dennis Calahan},
\author{Yifei Zhang},
\author{William M. Humphreys},
\author{Hyun Jung Kim*},
\author{Tian Gu},
\author{Juejun Hu*}





\begin{affiliations}
C. C. Popescu, B. Mills, Dr. L. Martin-Monier, L. Ranno, Prof. J. X. B. Sia, K. P. Dao, Dr. Y. Zhang, Dr. T. Gu, Prof. J. Hu\\ 
Department of Materials Science and Engineering, Massachusetts Institute of Technology, Cambridge, 02139, MA, USA\\

Prof. J. X. B. Sia\\
School of Electrical and Electronic Engineering, Nanyang Technological University, 50 Nanyang Avenue, Singapore 639798

Dr. K. Aryana, W. M. Humphreys, Prof. H. J. Kim, \\ 
NASA Langley Research Center, Hampton, 23666, VA, USA\\

Dr. T. W. Lee, Prof. H.-B. Bae\\ 
KAIST Analysis Center, Korea Advanced Institute of Science and Technology, Yuseong-gu, Daejeon 34141, Korea\\

B. Mills\\
Draper Scholar Program, The Charles Stark Draper Laboratory, Inc., Cambridge, MA 02139, USA\\

Dr. D. Calahan\\ 
The Charles Stark Draper Laboratory, Inc., Cambridge, MA 02139, USA\\

Dr. V. Liberman, Dr. S. Vitale\\ 
Lincoln Laboratory, Massachusetts Institute of Technology, Lexington, MA, 02421, USA\\

Prof. M. Kang,
New York State College of Ceramics, Alfred University, Alfred, NY, 14803, USA\\

Prof. K. A. Richardson\\ 
CREOL, The College of Optics \& Photonics University of Central Florida Orlando, FL, 32816, USA\\

Prof. C. A. R\'ios Ocampo\\ 
University of Maryland, Department of Materials Science \& Engineering, College Park, MD, USA\\

Dr. T. Gu, Prof. J. Hu\\ 
Materials Research Laboratory, Massachusetts Institute of Technology, Cambridge, 02139, MA, USA\\

Email Address: hujuejun@mit.edu, hj.kim67@gmail.com

\end{affiliations}


\keywords{chalcogenide glasses, phase change materials, metamaterials, on-chip photonics}

\begin{abstract}

Chalcogenide optical phase change materials (PCMs) have garnered significant interest for their growing applications in programmable photonics, optical analog computing, active metasurfaces, and beyond. Limited endurance or cycling lifetime is however increasingly becoming a bottleneck toward their practical deployment for these applications. To address this issue, we performed a systematic study elucidating the cycling failure mechanisms of Ge$_2$Sb$_2$Se$_4$Te (GSST), a common optical PCM tailored for infrared photonic applications, in an electrothermal switching configuration commensurate with their applications in on-chip photonic devices. We further propose a set of design rules building on insights into the failure mechanisms, and successfully implemented them to boost the endurance of the GSST device to over 67,000 cycles.

\end{abstract}

\section{Introduction}
Chalcogenide optical phase change materials (PCMs) claim giant refractive index contrast between their amorphous and crystalline states, a unique attribute that underlies their growing applications spanning reconfigurable photonic integrated circuits \cite{rios2022ultra,chen2023non,wuttig2017phase}, optical in-memory computing \cite{chakraborty2019photonic,chen2023neuromorphic}, nonvolatile displays \cite{carrillo2019nonvolatile}, as well as active metamaterials and metasurfaces \cite{fang2023non,zhang2021electrically}. 
For these applications, the ability to reversibly and reliably switch between the different structural states of PCMs is essential. Most studies on optical PCMs reported reversible switching between 10 to 10,000 cycles \cite{rios2022ultra, chen2023non, zhang2021electrically, martin2022endurance}. Endurance exceeding half a million cycles and several million cycles has recently been demonstrated in electrically switched waveguide-integrated PCM memories \cite{meng2023electrical} and via fast (20 kHz) laser switching \cite{lawson2024optical}. Impressive as these numbers are, they still fall short for many applications: a device switching at video frame rates (24 Hz) will hit one million cycles in just 11.6 hours! Understanding and mitigating the failure mechanisms that limit the cycling lifetime of PCMs are therefore of critical importance.

Before proceeding with further discussions on this topic, we want to point out that extensive studies have been carried out investigating failure mechanisms of chalcogenide PCMs in the context of phase-change random access memories (PCRAMs), and atomic migration caused by wind force, electrostatic force, and incongruent melting have been cited as the primary factors limiting cycling endurance in PCRAMs \cite{oh2020situ,yang2009atomic,kang2009analysis,kim2019phase}. Failure mechanisms of optical PCM devices, however, are expected to be different. Unlike electronic PCRAMs where the phase transition is triggered by passing electric current directly through the PCM, photonic devices incorporating PCMs resort to electrothermal switching using an external resistive micro-heater to prevent filamentation (a phenomenon where a thin wire of PCM first crystallizes, forming a low-resistance pathway that locally concentrates electric current precluding uniform switching of the entire PCM volume) \cite{you2015self,choi2010phase,sun2017nanoscale}. As a result, electric field-driven degradation mechanisms due to wind force and electrostatic force are insignificant in optical PCMs \cite{nam2012electrical}. Additionally, photonic applications typically involve a much larger PCM switching volume (of the order of 10$^{8}$ nm$^{3}$ in integrated photonic devices and $\mathtt{\sim}$ 10$^{14}$ nm$^{3}$ per 1 mm$^{2}$ aperture area for free-space devices \cite{kim2023p,gu2023reconfigurable}) compared to that in PCRAM ($\mathtt{\sim}$ 10$^{5}$ nm$^{3}$ or less). Consequently, thermal non-uniformity, chemical inhomogeneity, and mechanical stress are anticipated to play a far more significant role in degradation of PCM-based photonic devices. Lastly, the distinctive functional requirements for optical applications compared to PCRAMs have catalyzed the development of a wide variety of new PCM compositions, exemplified by low-loss PCMs such as Ge$_2$Sb$_2$Se$_4$Te (GSST) \cite{zhang2019broadband}, Sb$_2$S$_3$ \cite{dong2019wide} and Sb$_2$Se$_3$ \cite{delaney2020new}, whose failure mechanisms remain poorly understood.

Here we report a systematic study examining the failure mechanisms of GSST, a broadband transparent PCM that has enabled a wide spectrum of applications ranging from transient waveguide couplers \cite{zhang2021transient} to parfocal zoom metalenses \cite{shalaginov2021reconfigurable}. We investigated its switching behavior on a silicon-on-insulator (SOI) heater platform, which has been extensively adopted in PCM-based photonic devices given its compatibility with scalable Si foundry manufacturing \cite{rios2022ultra,chen2024low,sun2024microheater,zheng2020nonvolatile,erickson2023comparing}. In this comprehensive study, we evaluated the selection of encapsulation film material and thickness, assessed the impact of metal contact material and design, developed strategies to mitigate PCM delamination, and identified elemental migration due to incongruent melting as the main culprit of optical contrast reduction. Using an optimized design informed by insights into the failure mechanisms, we demonstrate reversible switching over 67,000 cycles, which is significantly improved from prior cycling endurance at $\mathtt{\sim}$ 1,000 measured in devices of similar configurations \cite{popescu2024electrically}.





\section{Uncovering failure mechanisms in optical PCMs}

The baseline device structure used to study the failure mechanisms is illustrated in SI Figure \ref{fig2:Fabrication}, which comprises a patterned PCM film resting on a doped SOI heater. The PCM film contains three types of patterns, 1-D line arrays, 2-D dot arrays, and unstructured patches to investigate the impact of PCM film morphology on endurance. Detailed device fabrication and measurement protocols are elaborated in Experimental Methods.

\subsection{Encapsulation layer}
We started out by investigating the impact of the PCM encapsulation layer. The encapsulation (capping) layer is necessary to protect the PCM against volatilization \cite{debunne2011evidence}, oxidation \cite{agati2020effects} and geometry distortion \cite{santala2014distinguishing} during cycling, a vigorous transient thermal process that involves melting the PCM. Several encapsulation materials have been applied to PCMs integrated in waveguides, such as ZnS-SiO$_2$ \cite{teo2023capping}, SiN$_x$ \cite{fang2021non} and Al$_2$O$_3$ \cite{zheng2020nonvolatile} typically with a thickness of 10 - 30 nm. Our initial attempt to use a 20 nm ALD Al$_2$O$_3$ layer as the encapsulation was unsuccessful. The problem arose because our device's PCM layer is much thicker (180 nm) compared to the thinner PCM films used in on-waveguide applications (around 20 nm). This significant thickness difference results in much greater stress from the volume change of the PCM during phase transitions (3\% for GSST \cite{shalaginov2021reconfigurable}). A marked increase of surface roughness was observed for a device with 10 nm Al$_2$O$_3$ on 45 nm of GSST after the first tens of cycles, and significant PCM material loss was observed as the device was further cycled, which we attribute to failure of the encapsulation causing partial PCM volatilization \cite{vitale2022phase}.

\begin{figure}[h!]
    \centering
    \includegraphics[width = 0.7\linewidth]{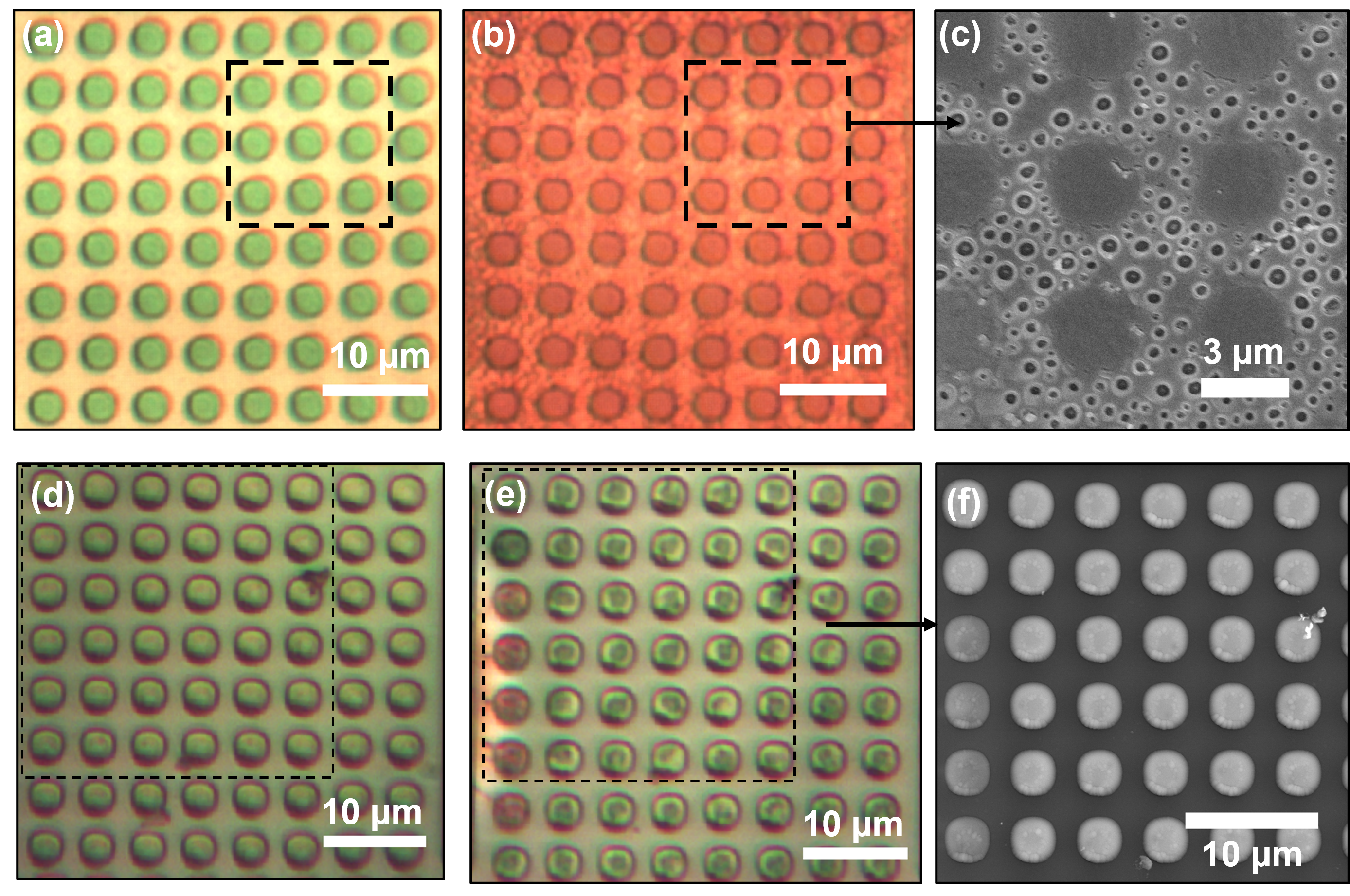}
    \caption{A device with 150 nm of GSST (green structures) and 400 nm of SiN$_x$ after (a) 2 cycles and (b) 900 cycles under optical microscope along with (c) SEM image of the device showing formation of pinhole structures linked to hydrogen evolution. Optical micrograph of a similar device with 180 nm GSST, 800 nm of reactively sputtered SiN$_x$ (d) before and (e) after 26,346 cycles, showing no noticeable changes in the optical behavior of the SiN$_x$. (f) SEM micrograph of the device with sputtered SiN$_x$ after 26,346 cycles, lacking the type of damage observed with PECVD SiN.}
    \label{fig3:HydrogenEvo}
\end{figure}

Since thick encapsulation layer deposition is impractical with ALD, we opted for bi-layer encapsulation with an additional SiN$_x$ on ALD Al$_2$O$_3$ to improve the durability of the encapsulation. Even though literature reported SiN$_x$ deposited by plasma enhanced chemical vapor deposition (PECVD) as a PCM capping material \cite{fang2021non}, we have identified hydrogen evolution in PECVD SiN$_x$ as a failure mechanism that compromises device longevity during cycling. In the experiment, 400 nm thick SiN$_x$ was deposited on the device using PECVD at 300 \celsius (substrate temperature). Upon cycling under consistent lightning, the SiN$_x$ layer started to change from clear and transparent (Figure  \ref{fig3:HydrogenEvo} a) to translucent with a reddish hue (Figure  \ref{fig3:HydrogenEvo} b) under an optical microscope. Upon closer examination using scanning electron microscopy (SEM), the SiN$_x$ layer was observed to have formed circular pinholes (Figure \ref{fig3:HydrogenEvo} c), which have varying sizes depending on their location on the heater (Figure  SI \ref{figSI:HydrogenMorphologyDevice}). This variation in size is likely due to a temperature gradient across the heater. Similar pinholes and damage have been previously observed in PECVD SiN$_x$ upon thermal treatment \cite{liu2011thermally,hughey2004massive,king2010intrinsic}, which was attributed to loss of hydrogen at elevated temperatures\cite{he1995hydrogen,yelundur2001enhanced}.

To resolve this issue, we turned to reactive sputtering of SiN$_x$ in a hydrogen-free (Ar/N$_2$) gas ambient. Figures  \ref{fig3:HydrogenEvo} (d) and (e,f) show a device encapsulated in 20 nm ALD Al$_2$O$_3$ and 800 nm sputtered SiN$_x$ before and after more than 26,000 applied cycles, respectively. The sputtered SiN$_x$ film remains defect-free in the process untill delamination (note the starting delamination at the left edge of the region) or contact failure occurs, as we shall discuss in following sections.

\subsection{Delamination}
\begin{figure}[h!]
\centering
\includegraphics[width = \linewidth]{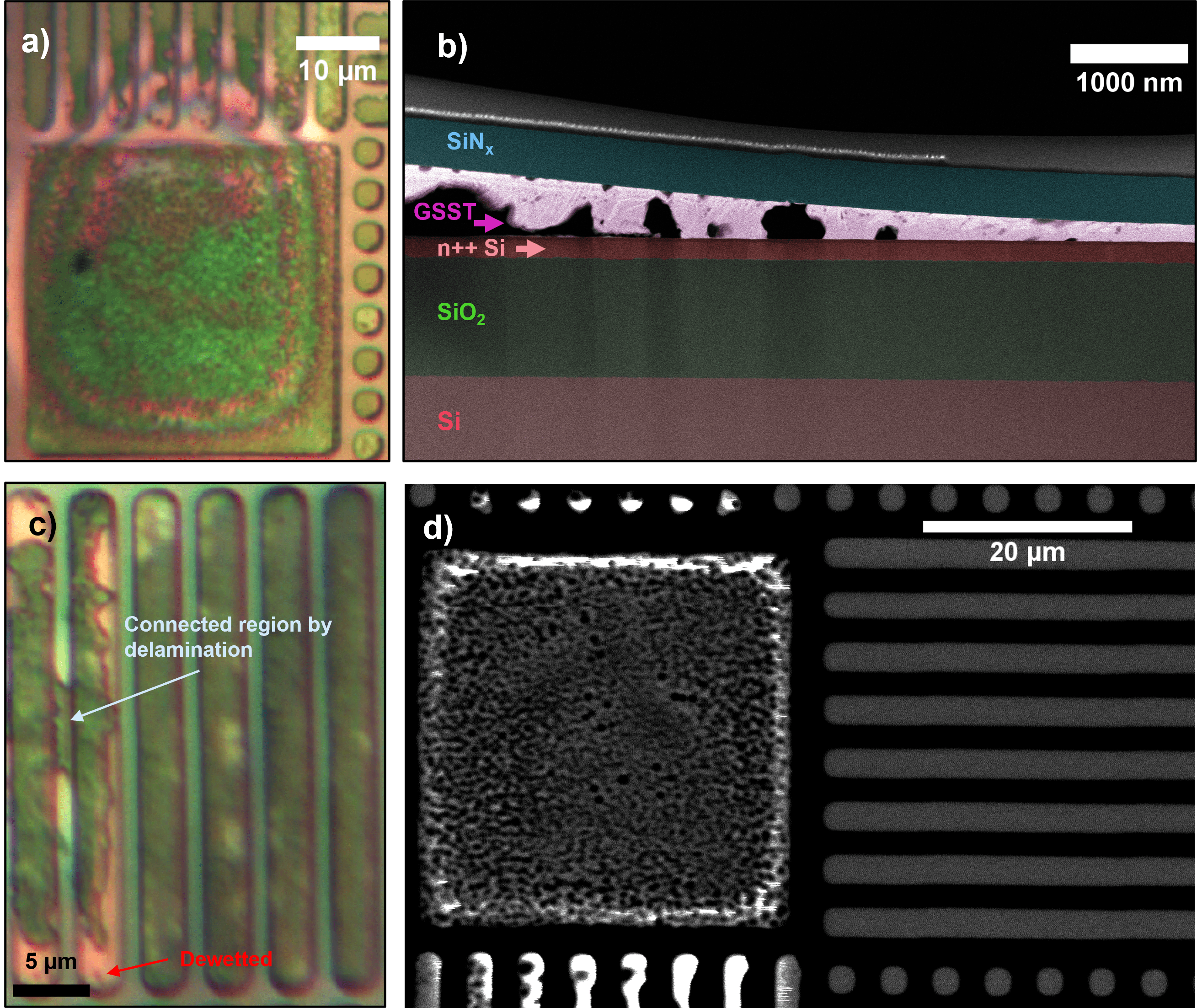}
  \caption{(a) Optical micrograph of a device showing the fringes typical of delamination on unpatterned PCM regions after spreading towards and being slowed by 1-D line structures. (b) A false color TEM image of the cross-section of a GSST film on a doped Si heater after delamination due to cycling, with the GSST showing preferential adherance to the Al$_2$O$_3$ / SiN$_x$ protective layer in comparison to the heater and (c) optical micrograph of GSST gratings showing the morphology of dewetted structures still in contact with the heater, with the pink-orange areas being the stack of nitride/aluminum oxide/heater (note the delaminated point where GSST leaked under the SiN$_x$ layer). (d) Backscattered SEM image of a cycled unpatterned region showing contrast variation from its 1-D line counterpart on the right due to mass flow, enabled by delamination of the top layer.}
  \label{figDel&Dewet}
\end{figure}

We observed two types of delamination failure in the PCM devices, delamination between the PCM and the underlying SiO$_2$-coated Si heater, and delamination between the Al$_2$O$_3$-SiN encapsulation layer and the heater. The former typically takes place within the unpatterned PCM patches starting at a few cycles to a few hundreds of cycles, and becomes readily visible under an optical microscope in the form of interference color fringes (Figure \ref{figDel&Dewet}a), indicating formation of pockets underneath the encapsulation layer. Cross-sectional transmission electron microscopy (TEM) inspection (Figure \ref{figDel&Dewet}b) reveals that the delamination preferentially occurs at the interface between the PCM and the heater. As a result, the PCM film lost thermal contact with the heater and stopped switching. The delaminated PCM film also dewets from the encapsulation layer, eventually forming isolated islands, which we will discuss further in the next section.

Such PCM-heater delamination can be effectively mitigated in patterned PCM structures (e.g., 1-D lines / gratings, 2-D dots, and fishnet-type metasurfaces \cite{popescu2024electrically}). Since ALD Al$_2$O$_3$ offers improved adhesion to SiO$_2$ {\cite{ding2011influence,takakura2023room}}, a patterned PCM structure provides anchoring points where the encapsulation layer directly contacts the heater, thereby preventing premature delamination. Alternatively, materials with improved adhesion to chalcogenide PCMs may lessen the impact of such delamination. Adhesion between PCM and another material can be gauged by their contact angle \cite{martin2021novel,gennes2004capillarity}, as a smaller contact angle implies lower interface energy and reduced tendency for delamination. The contact angles between a GeSbTe alloy and ZnS, ZnS-SiO$_2$ and SiO$_2$ have been reported as 82°, 89° and 128°, respectively  \cite{ebina2001wetGST-SiO2-ZnS}, suggesting that coating the Si heaters with ZnS or ZnS-SiO$_2$ prior to PCM deposition may suppress delamination. 

In patterned PCM structures, delayed failure can still occur due to delamination between the encapsulation layer and the heater at the anchoring points over thousands or tens of thousands of cycles (Figure \ref{figDel&Dewet}c). In patterned structures, we notice that the delamination most often initiates from the boundary between the patterned and unpatterned PCM regions (specifically from the unpatterned side)(Figure \ref{figDel&Dewet} a,d), and progressively propagates throughout the entire heater. Therefore, avoiding large unpatterned PCM regions is an effective way to delay the occurrence of these delamination events. Optimizing the Al$_2$O$_3$ ALD and SiN$_x$ sputtering deposition processes to enhance adhesion and lower compressive stresses in SiN$_x$ is another potential solution to further reduce the risk of delamination failure.

\subsection{PCM Dewetting}
Dewetting here refers to the retraction of PCM films resulting in uncovered heater areas. Figures \ref{figDel&Dewet} (a) and (c)  present two examples of dewetting occurring in an unpatterned PCM film and in a 1-D line pattern. Using high-energy electron back scattering which allows us to visualize elemental contrast with a penetration depth greater than the encapsulation layer thickness, we can clearly observe the retraction of PCM in Figures {\ref{figDel&Dewet}}c and {\ref{figDel&Dewet}}d, where the brighter regions correspond to GSST-covered areas. Given that amorphizing PCM involves a melt-quench process, we hypothesize that dewetting  takes place gradually during the time when the PCM is in a liquid phase, which is kinetically far more expeditious than solid-state dewetting. We note that failure linked to dewetting of PCMs in waveguide devices have also been identified in previous reports \cite{rios2022ultra}, suggesting that it is a common mode of failure in PCM-based photonic devices. We also note that dewetting is closely related to delamination, as the latter provides the excess volume needed for PCM retraction and void formation. For unpatterned PCM films, buckling of the encapsulation layer (visible in the form of color fringes Figure \ref{figDel&Dewet} a) is always concurrent with dewetting. Figure \ref{figDel&Dewet}c marks locations where delamination between the encapsulation layer and the heater happens, and the PCM film encroached underneath the encapsulation layer in these locations. In contrast, the other 1-D lines on the right side of the same image are free of delamination defects, and dewetting is correspondingly negligible.

Given the link between delamination and dewetting, measures taken to prevent delamination such as turning to patterned PCM structures are also effective in avoiding dewetting. Additionally, decreasing the volume of PCM also lowers the risk of dewetting. For example, empirically, we have found that the 2-D dots (2 to 4 $\mu$m in size) are practically immune to dewetting failure given the small, tightly confined PCM volume within each dot.

\subsection{Metal contact failure}
\begin{figure}[h!]
    \centering
    \includegraphics[width = \linewidth]{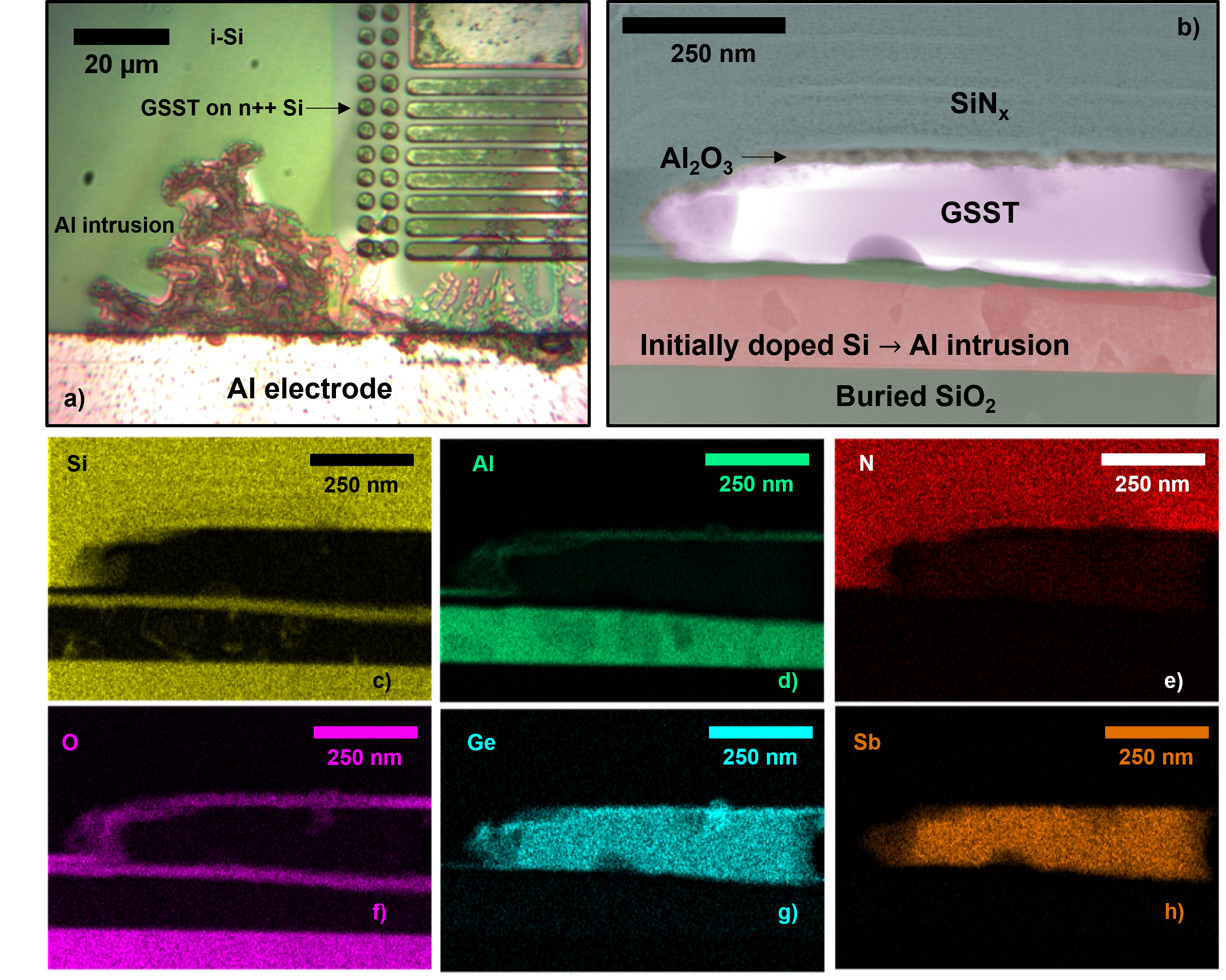}
    \caption{(a) Optical micrograph of device after failure via electromigration after 18,000+ cycles highlighting the typical dendrites forming at the edge of the heater (near the Al pad - doped Si - undoped Si point) and (b) a false color TEM image of a device after failure due to electrode failure, showing the propagation of the aluminum into the doped heater into a region with GSST. EDS Maps of corresponding elements from the TEM micrograph, highlighting that Si (c) was displaced in the original heater region by Al (d). The other elements, shown for reference, do not show intermixing. } 
    \label{fig2:ElectroMigration}
\end{figure}

Electrical shorting caused by the diffusion of the contact metal (Al in our case), along with delamination between the encapsulation layer and the heater, are two dominant failure mechanisms that limit the cycling endurance of our devices. Figures \ref{fig2:ElectroMigration}a and \ref{fig2:ElectroMigration}b show a top-view optical micrograph and a cross-sectional TEM image of a failed device due to Al diffusion, respectively. Elemental mapping (Figure \ref{fig2:ElectroMigration} c-h) indicates that Al has almost completely displaced Si in regions where such diffusion has taken place.

Our metallization process includes formation of a 10 nm Ti/ 20 nm TiN contact liner, which was intended as a diffusion barrier. The cause of failure of TiN barriers against Al diffusion is a subject that has been studied extensively in microelectronics \cite{harper1989mechanisms, kohlhase1989performance} and new barrier designs such as Ti/TiN multilayers have been proposed to mitigate the risk of TiN failure \cite{wu2005novel}. Alternatively, using metal contacts with enhanced robustness against thermal cycling, for instance, tungsten plugs in place of Al, will likely also improve the durability of the device.

\subsection{Elemental migration}
The mechanisms discussed in previous sections generally lead to catastrophic failure of the device abruptly, for example, due to film rupture, loss of thermal contact, or electrical shorting. In this section, we examine the origin of optical drift, herein referring to gradual (happening over thousands of cycles in our case) optical contrast reduction during GSST cycling, a phenomenon that has also been observed in other optical PCMs \cite{gao2021intermediate,xia2024ultrahigh,yuan2023electrode,chen2023non}.

\begin{figure}[h!]
    \centering
    \includegraphics[width = \linewidth]{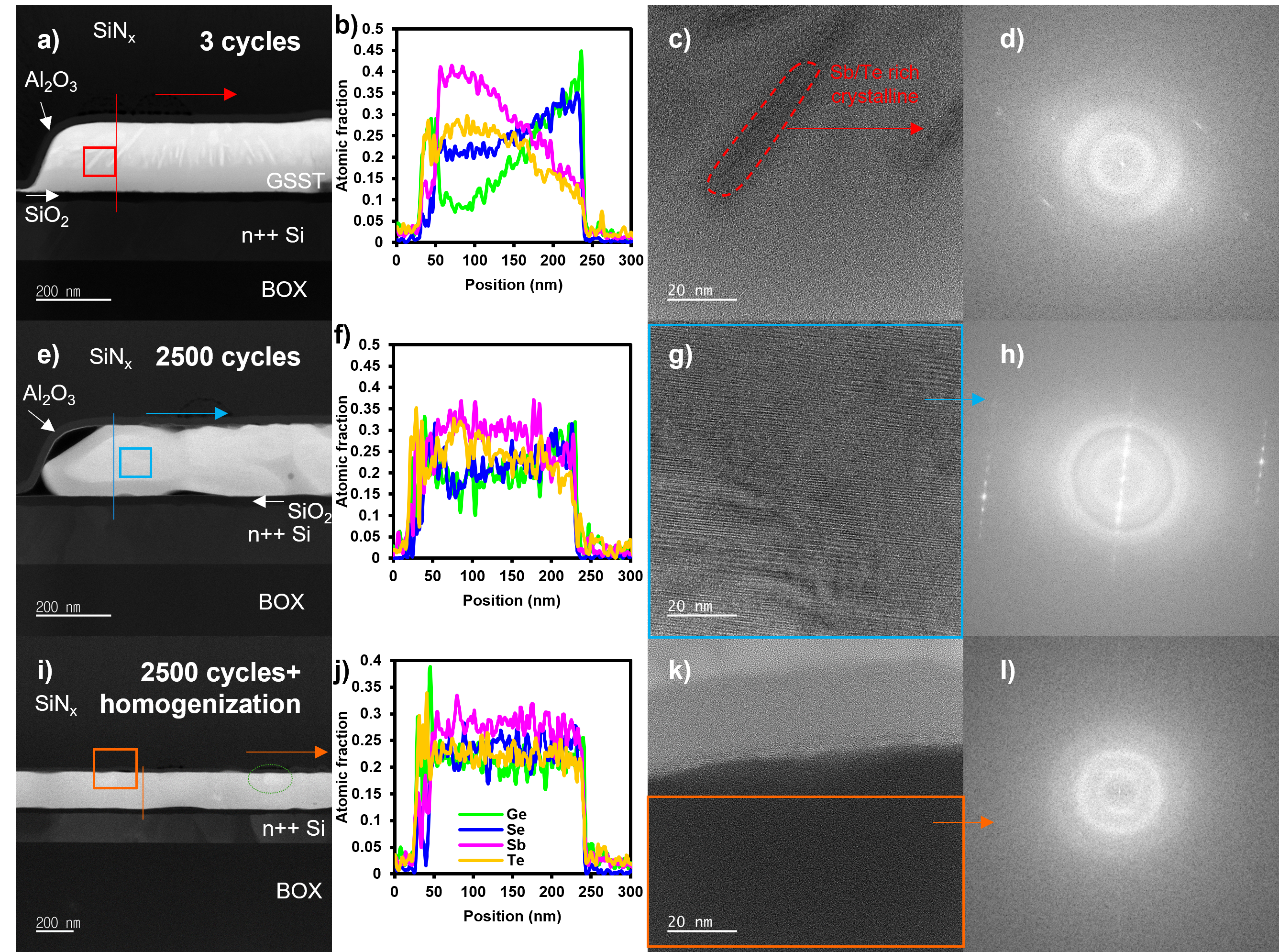}
    \caption{STEM images of the three samples taken from the end of a 1-D line near the electrode (a,e,i) with EDS line-scans across those regions from near the lines in the previous panels (b, f, j). HR-TEM images showing crystalline regions at the top of the film from (a) in (c), center of the image region from (e) in (g), and an amorphous region at the top of the film from (i) in (k). The approximate locations of the HR-TEM micrographs in the panels (a),(e) and (i) are highlighted in the corresponding color-coded squares.  Fourier transform images of the highlighted areas in (d),(h), and (l) showing that the top Sb+Te rich needle-like structure in (c) was still crystalline, the whole region in (g) was relatively crystalline and the top of the film, next to the Al$_2$O$_3$ layer in (k) was amorphized. While the SiO$_2$ and Al$_2$O$_3$ layers are captured in the images, for clarity, their positions are highlighted only in (a) and (e). The encircled area in (i) is one of the crystalline islands still present after homogenization.}
    \label{fig5:TEM}
\end{figure}

To start with, Figure \ref{fig5:TEM}a shows a cross-sectional scanning transmission electron microscopy (STEM) image of a PCM device (specially along a grating line in the 1-D line array pattern) after 3 switching cycles and upon application of a 33 V, 15 $\mu$s duration pulse to set it into the amorphous state. A composition gradient along its thickness direction is evident from the elemental contrast, where brighter areas are rich in heavier elements (Sb and Te) and vice versa, and also from the electron dispersive spectroscopy (EDS) line-scan result in Figure \ref{fig5:TEM}b. This composition gradient originates from incongruent vaporization of the constituent elements with different vapor pressures and has been well documented in other thermal evaporated chalcogenide films \cite{hu2007studies}. While the PCM is mostly amorphous, some needle-like crystals are visible in the Sb/Te-rich regions (Figures \ref{fig5:TEM}c and \ref{fig5:TEM}d).

\begin{figure}[h!]
    \centering
    \includegraphics[width = \linewidth]{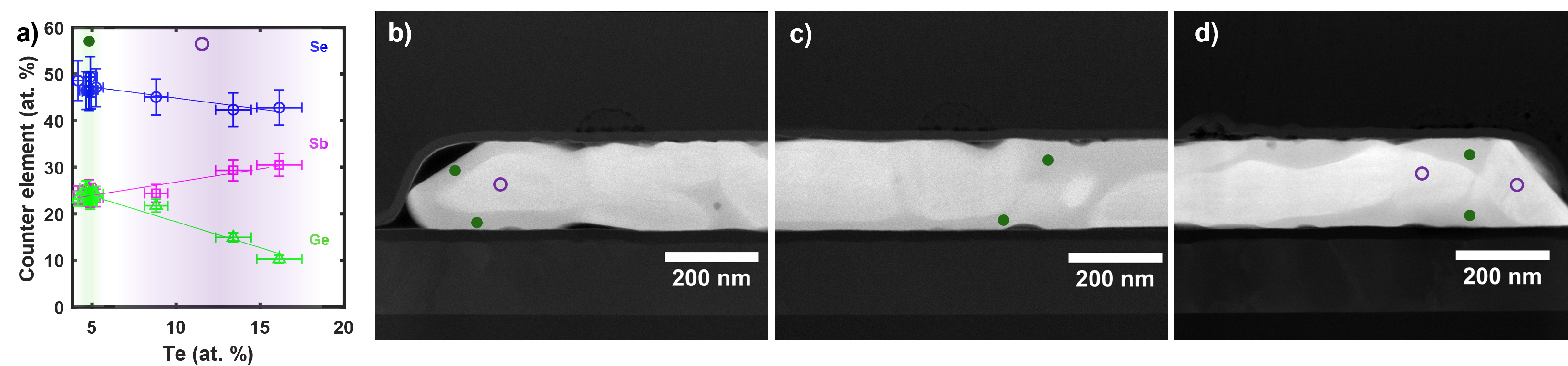}
    \caption{(a) Atomic concentration measured at various points for the partially compositional segregated sample for Ge, Sb and Se vs Te content (highlighting a region where the compositions cluster $\mathtt{\sim}$5 \% Te and three outlier points corresponding to higher intensity regions in STEM. (b-d) Left, middle and right side of the sample highlighting that the green symbols correspond to a lower intensity regions and the other three points corresponding to higher intensity regions in the STEM images. Ge and Se decrease with increasing Te content while Sb increases.}
    \label{fig6:PhaseNodeMosaic}
\end{figure}

In comparison, Figure \ref{fig5:TEM}e shows a PCM device on the same chip, with an identical configuration, and similarly upon being subjected to a 33 V, 15 $\mu$s duration pulse albeit after 2,500 switching cycles. Two striking differences have taken place compared to the sample in Figure \ref{fig5:TEM}a. The through-thickness composition gradient has been removed (also see Figure SI \ref{figSI:Sb_ThreeLayers_TEM}), which we attribute to liquid-phase inter-mixing and homogenization during repeated melting-amorphization cycles \cite{xie2018self}. This type of gradient removal occurs relatively early in the lifetime of a device, as can be seen when a deliberate large gradient was introduced via a triple deposition step. After 100 cycles, the initial profile cannot be observed anymore (Figure SI \ref{figSI:Sb_ThreeLayers_TEM}).  The elimination of the through-thickness composition gradient also explains the burn-in behavior observed in our PCM devices, where large optical contrast fluctuations occur within the first few tens of cycles before the devices settle into a more stable state without drastic cycle-to-cycle changes. In place of a consistent composition variation trend in the out-of-plane direction, the material evolves into two inter-dispersed phases that are randomly distributed (Figure {\ref{fig6:PhaseNodeMosaic}} b-d). The phase rich in Ge and Se is amorphized whereas the phase rich in Sb and Te stays mostly crystalline as shown by the high-resolution TEM (HR-TEM) image in Figure \ref{fig5:TEM}g. The inability to completely amorphize the PCM leads to gradual reduction of the effective switching volume and the observed optical drift.

To better understand the cause of this partial amorphization phenomenon, we applied an additional 33 V, 30 $\mu$s duration pulse on another otherwise nominally identical sample to homogenize it. Figure \ref{fig5:TEM}i shows that this more aggressive pulse amorphizes most of the PCM with only a few scattered small crystals left (see also Figure \ref{figSI:Island_After_Homogenization}). This finding implies that the Sb/Te-rich phase is more stable than the Sb/Te-rich crystals during initial cycles (Figure \ref{fig5:TEM}a), and thus its higher liquidus temperature prevents complete melting during the normal amorphization cycle (using 33 V, 15 $\mu$s duration pulses). Similar elemental migration phenomena have also been investigated in PCRAMs and two mechanisms have been proposed: thermal migration and incongruent melting \cite{park2007phase,yang2009atomic,yang2013driving,kang2009analysis}. Our finite-element method (FEM) simulations show that our PCM device experiences only a minor temperature gradient in the thickness direction with a maximum temperature difference of 1 K (Figure SI \ref{figSI:COMSOL}). Even if some thermal migration is driven by this temperature difference, the resulting composition gradient should also be along the thickness direction, which contradicts the random phase distribution observed in our experiment (Figure \ref{fig6:PhaseNodeMosaic} b-d). Therefore, incongruent melting, i.e., segregation of a single solid phase into one solid and one liquid phase at elevated temperatures, is likely a plausible explanation for the elemental segregation. While a phase diagram for GSST is unavailable to fully ascertain this hypothesis, it is well documented that Ge$_2$Sb$_2$Te$_5$, a close analog of GSST, does suffer from incongruent melting \cite{shelimova2000structural}. The negative correlation between Ge and Sb has been noted before for Ge$_2$Sb$_2$Te$_5$ \cite{kang2009analysis}. Using PCMs having a single-phase region extended to the melting point is a potential way to eliminate elemental segregation due to incongruent melting. Several antimony-based binary optical PCMs such as Sb$_3$Te$_7$, Sb$_2$S$_3$, Sb$_2$Se$_3$ meet this criterion, although their non-cubic crystalline structures can incur excess scattering optical losses \cite{li2024performance}.

The elimination of compositional non-uniformity by the aggressive amorphization pulse (Figure \ref{fig5:TEM}i) corroborates our conclusion that melting and liquid-phase transport is an effective means to compositionally homogenize PCM and suppress optical drift. However, excessively aggressive amorphization pulses also tend to expedite delamination between the encapsulation layer and the heater. Therefore, dynamically optimizing the electrical pulse parameters is important to maintaining a consistent optical contrast in PCM devices. After a homogenization pulse, most of the crystalline structure is erased, necessitating a more extensive crystallization process to achieve a fully crystallized state throughout the material. In this case, the crystalline islands at the edge of the device can prove advantageous (SI Figure \ref{figSI:Island_After_Homogenization}). The optical contrast is temporarily recovered, and the switching area increases with each cycle. This is because, as it has been seen before (Figure \ref{fig5:TEM} a-h), the typical 15 $\mathrm{\mu}$s pulses do not fully remove the crystalline structure, providing also a slight distinction between the fully amorphous and partially amorphous regions under optical microscope (SI Figure \ref{figSI:Homogenization_Optical} (a-c)). Although further analysis of the optical contrast over thousands of cycles after homogenization was not conducted, it is likely that the material will gradually revert to its phase-segregated state due to the incomplete amorphization and incongruent melting, requiring reset/homogenization pulses periodically.


\section{Optimizing PCM devices toward high endurance}

In this section, we discuss our efforts aiming to enhance the endurance of PCM devices building on the insights on their failure mechanisms. The optimized device under test is encapsulated with bi-layer ALD Al$_2$O$_3$ (20 nm) / sputtered SiN$_x$ (800 nm) and assumes a 2-D dot array geometry to minimize risks of delamination and PCM dewetting. To counter optical drift, the electrical pulse parameters were manually adjusted over the cycles, in small voltage and time increments. Care was taken to avoid overly aggressive amorphization pulses to prevent premature delamination failure. Though not implemented in this study, we note that our recently developed computer algorithm \cite{garud2024robust} provides a facile route enabling automated, dynamic adjustment of the pulse parameters without human intervention, thereby boosting consistency and endurance of optical PCM devices. Constrained by compatibility with our foundry process, the device still adopted Al metal contacts with Ti/TiN liners.

\begin{figure}[h!]
    \centering
    \includegraphics[width = \linewidth]{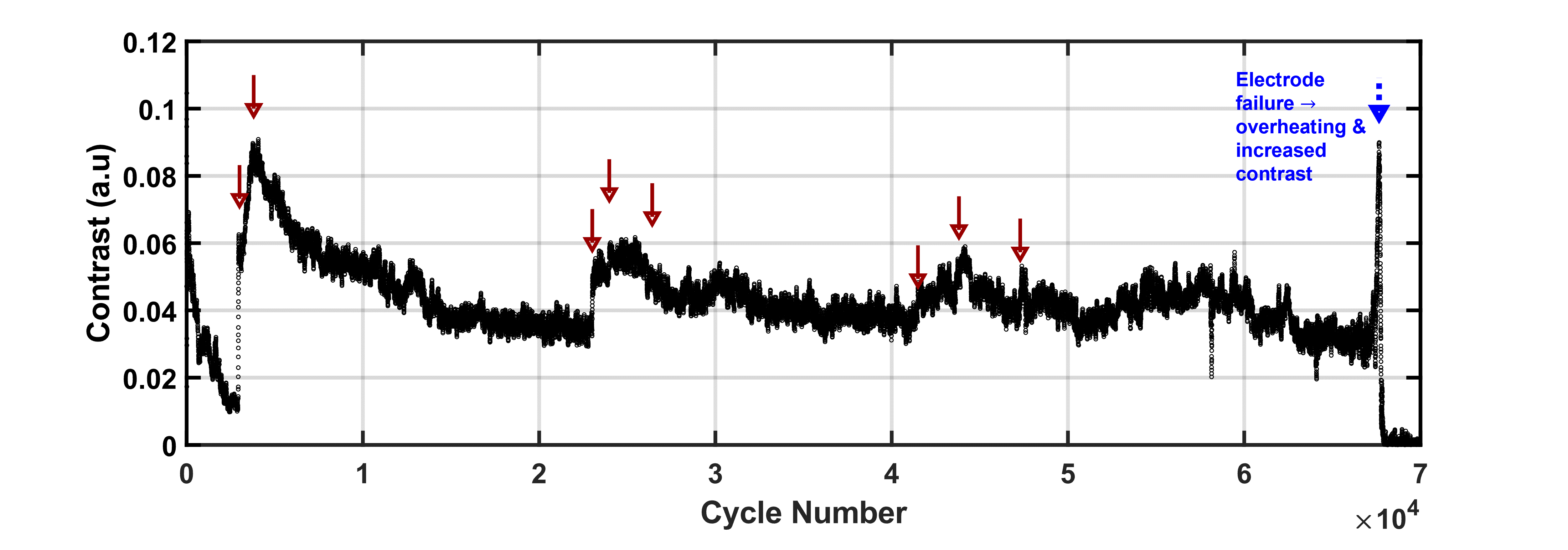}
    \caption{Optical reflectance contrast in a 150 $\mu$m × 150 $\mu$m PCM device. The contrast was smoothed with a moving average of 20 points for a better illustration of the significant trends. The solid arrows highlight the cycles where adjustments of the voltage or time of applied pulses were manually made, with a gradual cycle-to-cycle increase between about 2900 and 3800 for potential damage prevention. The dashed arrow highlights the point where progressive electrode damage leads to local overheating and a transient increase in the contrast before the catastrophic electrode failure takes place.}
    \label{fig1:Contrast}
\end{figure}

Figure \ref{fig1:Contrast} plots the measured optical reflection contrast of the device over 70,000 cycles. Time points where the pulse parameters were manually adjusted are labeled with arrows. A subset of images of the analyzed region has been collated into a video available in Supporting Information. Better consistency can be obtained by turning to computer algorithm assisted pulse optimization instead of human intervention \cite{garud2024robust}. The device maintained reversible switching with a large optical contrast up to $\mathtt{\sim}$ 67,500 cycles, when electrical shorting due to Al contact failure took place. This result marks a major improvement over our prior record of 1,250 cycles, and represents the highest endurance reported in large-area PCM photonic devices\cite{popescu2024electrically}. Even though higher endurance values have been achieved in small-area PCM structures \cite{meng2023electrical,lawson2024optical}, increasing switching volume (in our case over $10^{4}$ times larger than these previous reports) presents significantly elevated risks for all the aforementioned failure mechanisms given the much larger stress, applied electric current/voltage, thermal and structural non-uniformity, and probability of structural defect initiation. In addition, it is also noted that the endurance of our device is presently limited by failure of the Al metal contacts rather than the PCM itself. Further improvement is therefore anticipated by resorting to more robust metal contact designs.

\section{Discussion and Conclusion}

In this study, we have identified five failure/degradation mechanisms limiting the endurance of electrothermally - switched optical PCM devices and discussed/implemented their respective mitigation strategies outlined below.

\textbf{- Mechanical/chemical failure of encapsulation layers:} we adopted Al$_2$O$_3$-SiN bilayer encapsulation to furnish adequate mechanical robustness against cyclic stress due to PCM volume change upon phase transition. Chemically unstable films at elevated temperatures, such as PECVD SiN, should be avoided.

\textbf{- Delamination:} Patterned PCM structures, which provide anchoring points where the encapsulation layers directly contact the heater with improved adhesion, are preferred over unpatterned PCM films. Introduction of adhesion-promoting layers, optimization of the encapsulation layer deposition process to lower internal stress, and improving heater temperature uniformity to eliminate local "hot spots" can also suppress delamination.

\textbf{- Dewetting:} Measures used to circumvent delamination also apply to minimizing dewetting. Using structures with a small confined PCM volume (e.g., isolated PCM meta-atoms) lowers the risk of dewetting.

\textbf{- Metal contact failure:} Metal diffusion and resulting shorting can be mitigated with diffusion barriers and the use of refractory metals as contacts. Optimized heater and metal contact designs to reduce the temperatures at the contacts and possible current crowding are also helpful.

\textbf{- Elemental segregation:} Using a PCM that remains single phase continuously up to its melting point avoids elemental segregation due to incongruent melting. Adjusting the amorphization voltage pulse parameters to ensure complete melting of PCM and sufficient inter-mixing facilitated by liquid phase transport is a viable means to reverse elemental segregation and the resulting optical drift.

A 150 $\mu$m $\times$ 150 $\mu$m PCM device was implemented following these recommendations and achieved an endurance of over 67,000 cycles, representing a significant improvement over previous demonstrations of large-area PCM optical devices. This number is at present limited by the unoptimized metal contact design, suggesting that much higher endurance approaching the fundamental material limit is possible in the future. Results from our study inform important guidelines for optimization of next-generation PCM-based optical devices toward enhanced reliability, and can empower exciting new applications ranging from micro-display to dynamic beam shaping where high endurance is essential.

\section{Experimental Methods}

\subsection{Fabrication}
\begin{figure}[hb!]
    \centering
    \includegraphics[width = \linewidth]{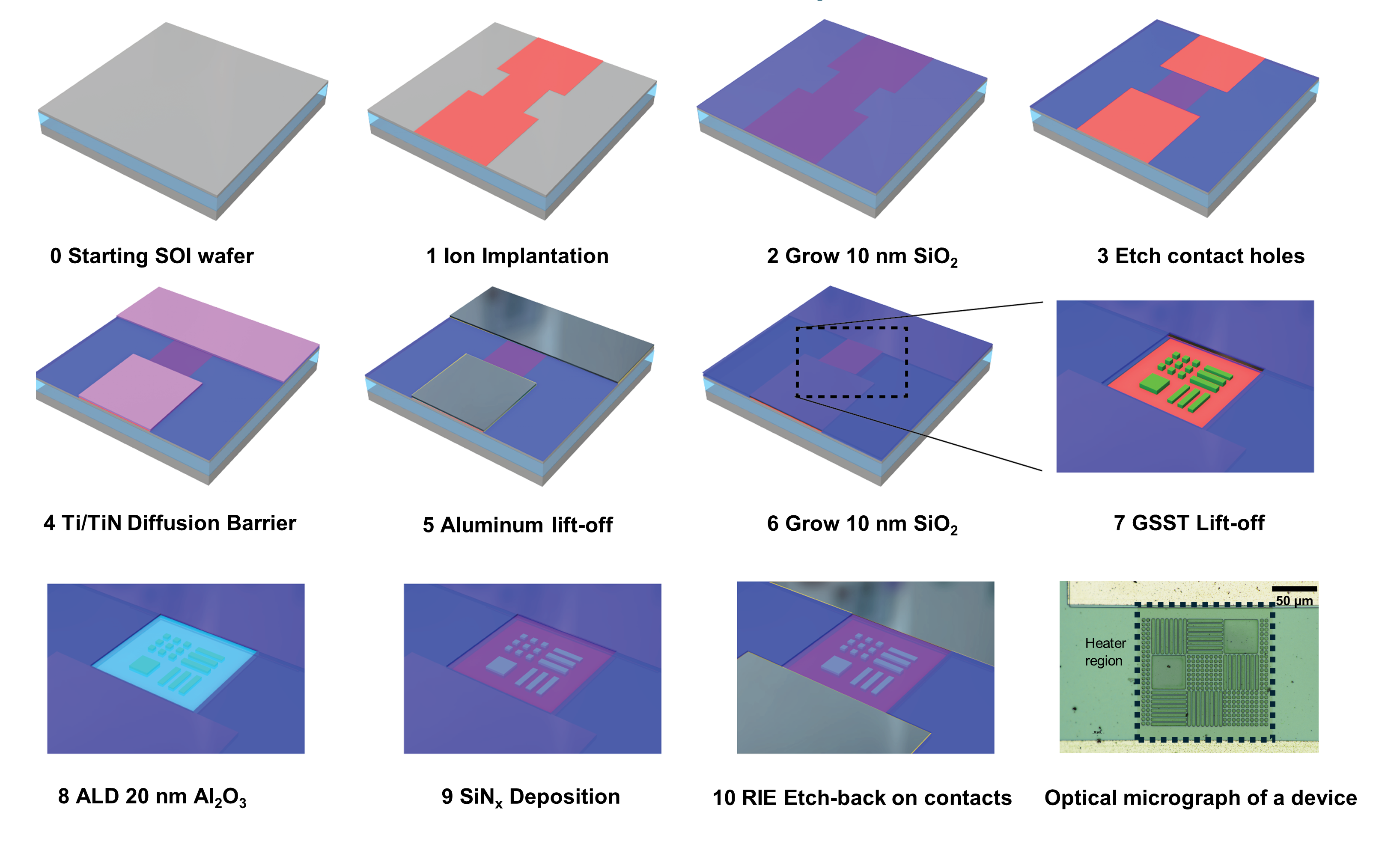}
    \caption{ Schematic of fabrication steps from the starting SOI wafer, ion implantation and metallization (performed at Lincoln Laboratories) followed by the PCM lift-off, ALD encapsulation, SiN$_x$ deposition (PECVD or reactive sputtering) and etch-back. The micrograph shows a 150 $\mu$m tested device.The heater region is highlighted via the dashed square line.}
    \label{fig2:Fabrication}
\end{figure}

\begin{figure}[h!]
    \centering
    \includegraphics[width = \linewidth]{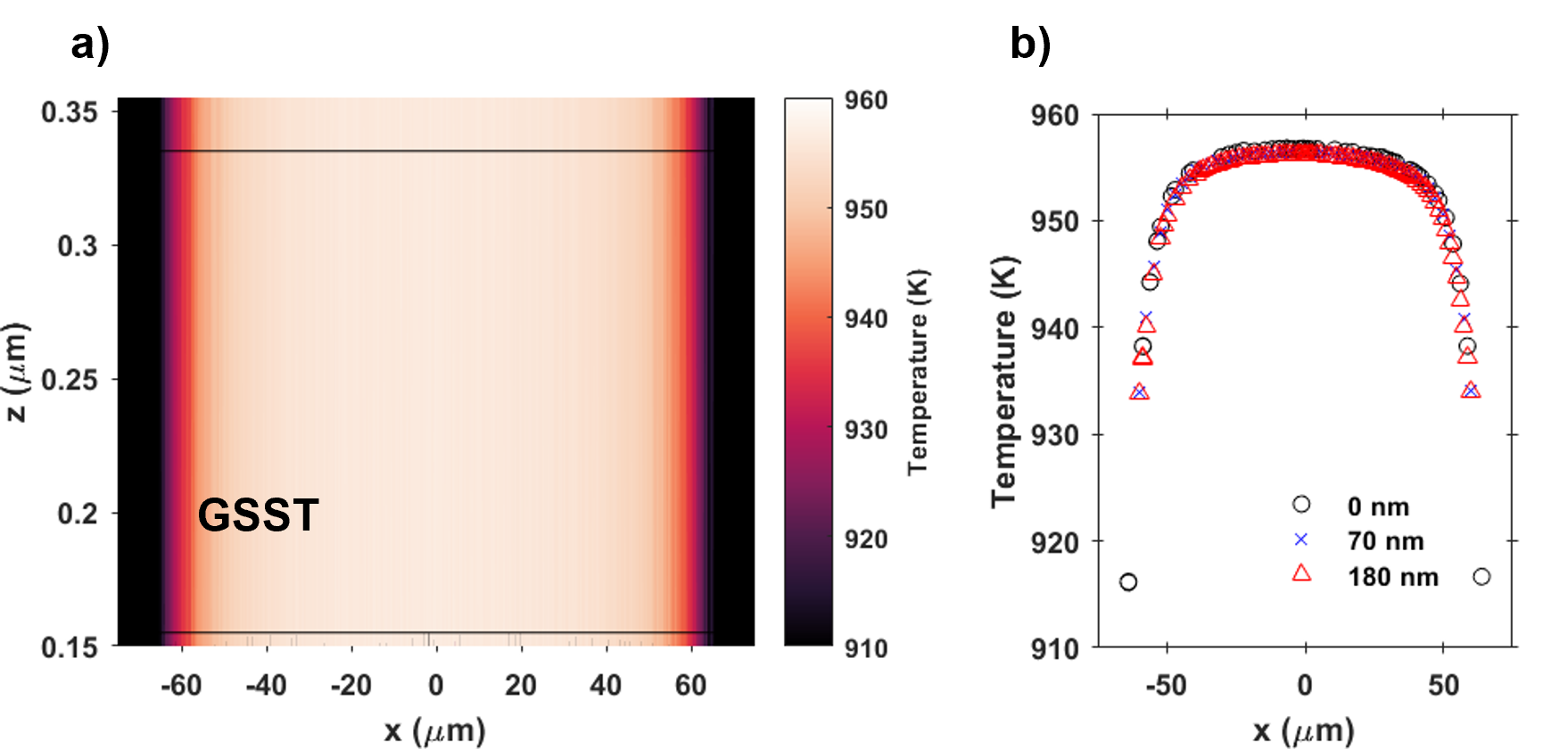}
    \caption{(a) A temperature map in cross-section through a thin film GSST (180 nm) on doped Si heater (155 nm) with SiN$_x$ (800 nm) capping layer from COMSOL simulations. Large variations are observed primarily at the boundary. The black lines delineate the GSST film from the structures encasing it. (b) Temperature of the GSST film in the cross-section at selected heights from the GSST-heater interface. Due to the ability of heat to dissipate both through the nitride layer and into the Si substrate below the 1 $\mu$m of buried oxide, the temperature gradient along the thickness of the structure is virtually non-existent. }
    \label{figSI:COMSOL}
\end{figure}

The SOI micro-heater platform was fabricated in the Lincoln Laboratory Microelectronics Laboratory \cite{popescu2023open}. Specific steps that point at important design considerations for micro-heaters for PCM photonics are highlighted below. In short, starting with 200 mm SOI wafers (150-160 nm of Si on $\mathtt{\sim}$ 1 ${\mu}$m of buried oxide), a heavy ion implantation step was performed with P at 80 keV with $\mathtt{\sim}$ 10$^{16}$ cm$^{-2}$ dose (with an estimated maximum doping concentration of $\mathtt{\sim}$ 5 $\times$ 10$^{20}$ cm$^{-3}$), followed by a rapid thermal annealing (RTA) step for 10 s at 1000 \celsius. 10 nm of SiO$_2$ were grown on the heater, followed by the opening of contact holes above the regions for electrical contact for each individual heater/device. 10 nm of Ti (adhesion layer) and 20 nm of TiN (diffusion barrier) were grown on top of the chip, followed by a 700 \celsius \ rapid thermal annealing step. The liner layer was patterned and etched to cover all the regions that the subsequent aluminum layer was deposited on. An aluminum electrode layer was deposited via lift-off, before an extra 10 nm of SiO$_2$ was deposited for protection of the devices. Afterwards, AZ nLOF 2020 was patterned using a laser direct writer (Heidelberg MLA 150) followed by development in AZ 300 MIF as a lift-off mask. Within each micro-heater, the PCM layout contained unpatterned patches of films, 1-D periodic lines at 5 $\mu$m pitch, and 2-D dots in a square lattice also at 5 $\mu$m pitch. Line patterns parallel and perpendicular to the applied electric field were both included. GSST was then deposited via thermal evaporation from GSST powder placed in a R.D. Mathis Ta baffled boat at a base pressure of $\mathtt{\sim}$ 10$^{-6}$ Torr. A relatively large GSST thickness of $\mathtt{\sim}$ 180 nm was chosen in this study, which resembles the thickness used in PCM-based metasurfaces \cite{zhang2021electrically}. After the GSST deposition, the AZ nLOF resist was removed via overnight immersion in n-methylpyrrolidone (NMP), before it was rinsed in acetone and then in isopropyl alcohol (IPA). Afterwards, the sample was encapsulated in 20 nm of Al$_2$O$_3$ via ALD at 150 \celsius \ (Savannah Thermal ALD). For the optimized device, a 800 nm SiN$_x$ layer was subsequently coated via reactive sputtering in an AJA ATC-Orion 5 sputterer using two 2-inch silicon targets at a pressure of 3 mTorr with Ar:N$_2$ 1:1 gas flow ratio, 12 sccm total flow rate, and 100 W RF power on each target. The PECVD SiN$_x$ films were deposited at 300 \celsius \ in a STS PECVD via a standard mixed frequency deposition process. The etch back to the metal contacts was done via patterning and SF$_6$ based reactive ion etching. We intentionally allowed additional etching time such that the Al$_2$O$_3$ ALD layer was also etched to facilitate wire-bonding. After resist removal in oxygen plasma (250 \celsius \ in an ESI 3511V-001 asher), the sample is mounted and wire-bonded with gold wires in a 1204 MEI ball bonder to a custom-made PCB (design available from prior work \cite{popescu2023open}). A schematic of fabrication process can be found in Figure SI \ref{fig2:Fabrication}.

\subsection{Characterization}

The devices were crystallized using voltage pulses with peak amplitudes of 18 - 20 V and durations of 15  - 60 ms. For amorphization, pulse amplitudes between 32.6 - 35 V and durations between 13 - 15.7 $\mu$s were used. The optical micrographs taken during cycling were collected with an AF 205 AmScope autofocus camera with a 40x objective. The contrast values were obtained based on the method described in our prior work \cite{popescu2023open} using a threshold value of 8 across all three color channels on an 8 bit-depth range. The contrast value is defied as $C = \frac{\overline{{I_{PCM}}_{cr}}-\overline{{I_{PCM}}_{am}}}{\overline{{I_{PCM}}_{cr}}}$ with $\overline{{I_{PCM}}_{cr}}$ and $\overline{{I_{PCM}}_{am}}$ being the average of the pixel count in the regions identified as switching(i.e. with PCM) for the crystalline (cr) and amorphous (am) state. The transmission electron microscope (TEM) samples were prepared in a Thermo Fisher Scientific focused ion beam system Helios G5 UX,  and analyzed in a Cs-corrected TEM (Titan cubed G2 60-300, Thermo Fisher Scientific ) at 300 kV. The FTIR spectra were collected on an iS50 Nicolet micro-FTIR.

\medskip
\textbf{Acknowledgements} \par 
Funding support to this work is provided by NSF under Awards 2132929, 2329088, DMR-2329087/2329088, supported in part by funds from federal agency and industry partners as specified in the Future of Semiconductors (FuSe) program, and by the Air Force Office of Scientific Research (AFOSR) under award number FA9550-22-1-0532. C.C.P would like to acknowledge insightful discussions with Prof. Carl V. Thompson on the mechanisms discussed in this manuscript. This work was carried out in part through the use of MIT.nano's and Harvard Center for Nanoscale Systems (CNS) facilities. Special thanks are extended to Mr. Ronald Neale for his contributions to graphic design. B. Mills acknowledges support provided by the Draper Scholar Program.

\medskip

%
\bibliographystyle{MSP}
\bibliography{ref}


\medskip
\textbf{Supporting Information} \par 
Supporting Information is available from the Wiley Online Library or from the author.

\section{Supplementary information}


\textbf{Fabrication schematic} \newline

\textbf{Thermal simulations} \newline
A COMSOL simulation for a transient electric pulse typical of the amorphization process was performed (32 V, 15 $\mu$s, accounting for lost power due to series resistance). The assumed system had a 150 $\mu$m heater with Si substrate, 1 $\mu$m of buried oxide, 155 nm of doped Si, 180 nm of GSST and around 800 nm of SiN$_x$ from the top of the GSST. The GSST film was assumed to be a square of 120 $\mu$m, covering most of the heater area. The 20 nm of SiO$_2$ on the heater and 20 nm of Al$_2$O$_3$ on top of the GSST were not part of the simulation due the very high refinement needed for the mesh to incorporate them and their similar thermal conductivity (in order of magnitude) to the nitride and GSST, resulting in a low likelihood of their presence altering the results significantly. More information about the simulation setup can be found elsewhere \cite{popescu2023open,popescu2024electrically}. The snap-shot of the cross-section shows no significant temperature gradients across the film appears, and the main gradients of interest occur in-plane. If heat was unable to escape through the buried oxide, such as in a scenario with virtually zero thermal conductivity or an infinite insulating layer, a large thermal gradient would develop, indicating unidirectional heat diffusion. In other words, the thermal gradient shows the direction of heat flow so large thermal gradients show that heat flows in a particular direction. In this system, however, only 1 $\mu$m of buried SiO$_2$ is present and the 800 nm of SiN$_x$ is assumed to have around 2 W m$^{-1}$ K$^{-1}$, allowing for a bidirectional heat flow from the point of the GSST film.  

\begin{figure}[h!]
    \centering
    \includegraphics[width = 0.5\linewidth]{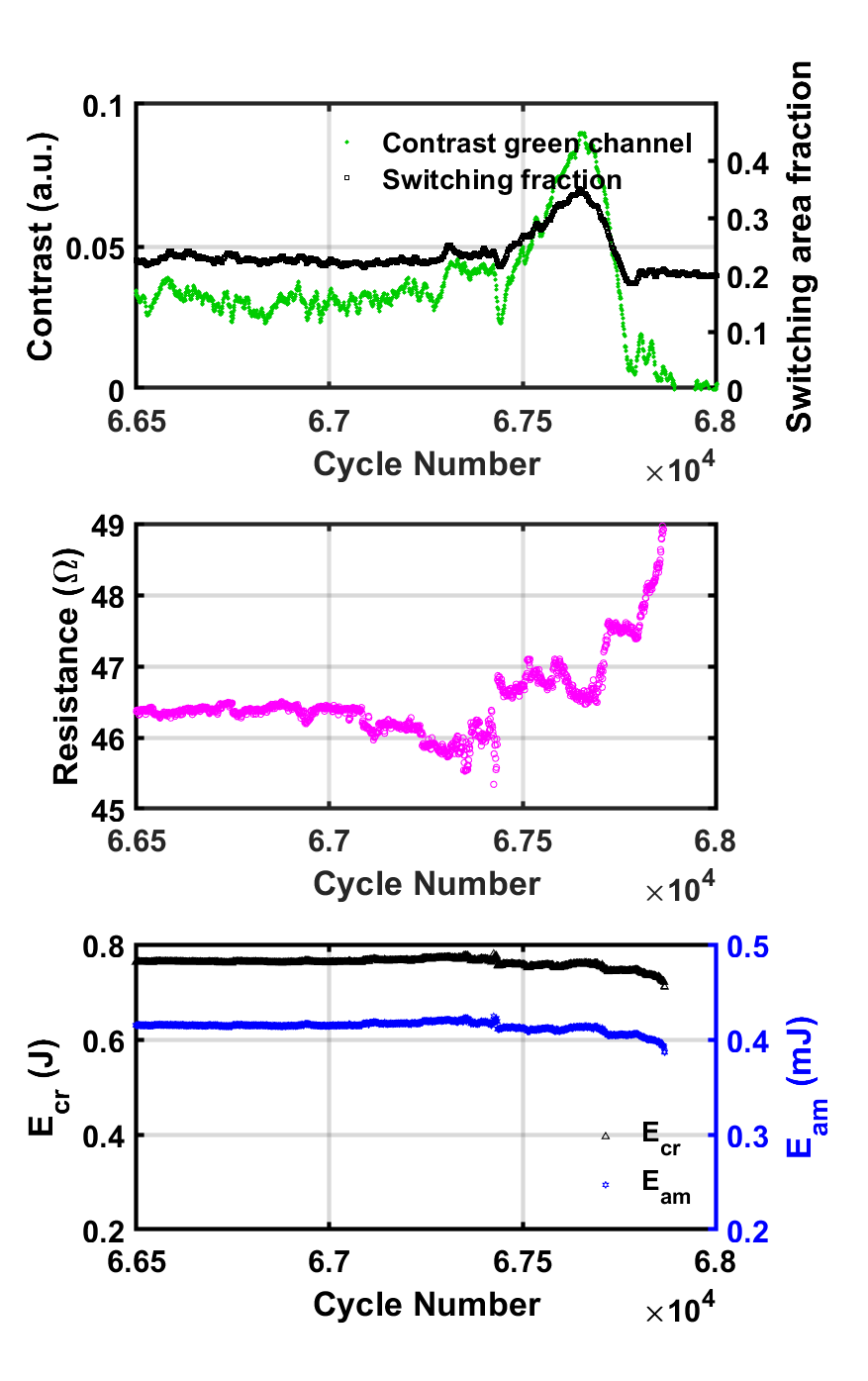}
    \caption{The contrast and switching area fraction values (a), resistance (b) and applied energy during amorphization and crystallization (c) of the device from  Figure  \ref{fig1:Contrast} highlighting the end of the life of this device before electrode failure.}  
    \label{figSI:resistance}
\end{figure}

\textbf{Device behavior at end-of-life} \newline
Close to the end of the lifetime of the device, aluminum infiltration into the doped Si region causes local overheating and thus the GSST contrast window is increased temporarily. The effects of these can be seen in an increased contrast, increased switching area fraction and significant changes in the resistance (Figure Si \ref{figSI:resistance}). While there were a few downward spikes in the resistance, it had an upper trend close to failure. This is likely due to the multi-part process of spiking of the Al into the Si layer \cite{chang1988aluminum,cerva1998aluminum}.

\begin{figure}[h!]
    \centering
    \includegraphics[width = 0.5\linewidth]{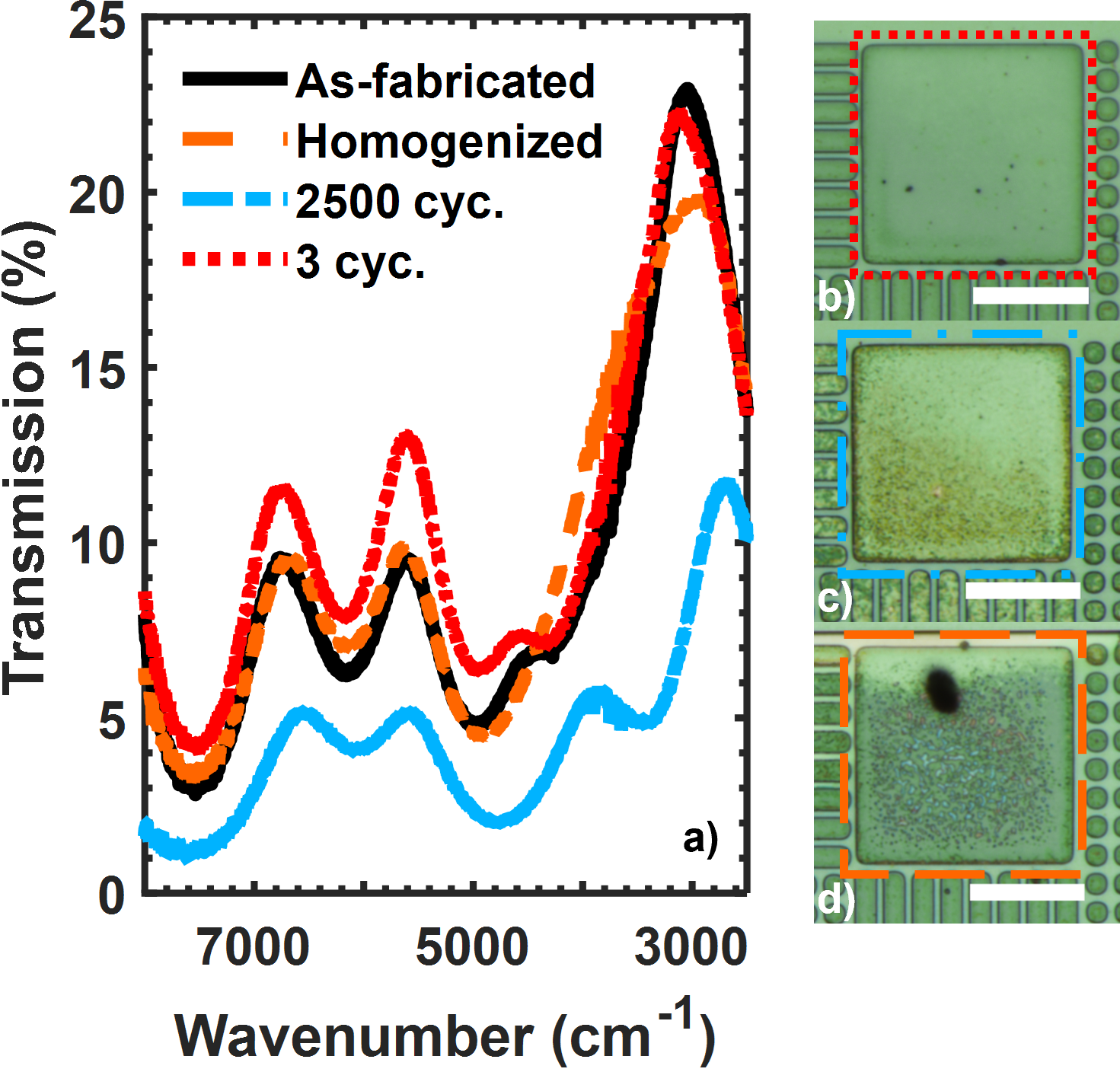}
    \caption{(a) FTIR transmission of the thin film from the 150 $\mu$m devices analyzed in TEM with optical micrographs of the specific regions from where the transmission values were collected. The sample was not backside polished. The scale bars are 20 $\mu$m.}  
    \label{fig:FTIR_TEM}
\end{figure}

\textbf{IR Transmission comparison} \newline
Micro-FTIR measurements on the devices used for elemental migration analysis show that the homogenization pulse is able to recover similar transmission levels for the device as it had at the beginning of its testing and comparable to an as fabricated device Figure SI \ref{fig:FTIR_TEM}. The transmission measured agrees also with the expected optical behavior that would be inferred from the TEM results. 

\begin{figure}[h!]
    \centering
    \includegraphics[width = \linewidth]{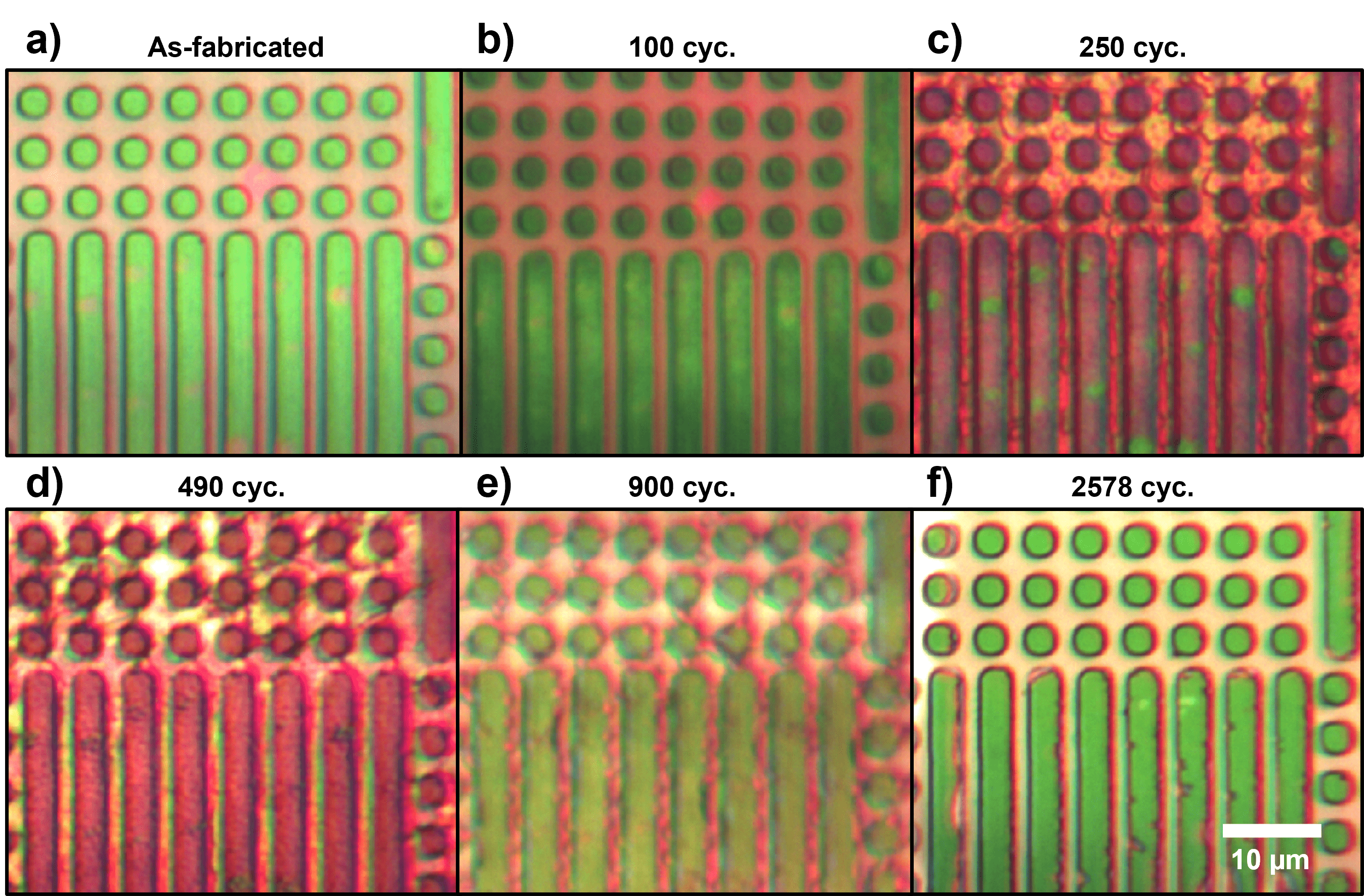}
    \caption{Optical micrographs of a device (150 nm GSST, PECVD SiN$_x$) as it was cycled using 36 V 10 $\mu$s pulses for amorphization and 16 V, 1 s for crystallization. The reversal of the optical behavior due to the hydrogen evolution can be observed over the course of the testing.}  
    \label{figSI:Opt_Hydro_Evol}
\end{figure}

\begin{figure}[h!]
    \centering
    \includegraphics[width = \linewidth]{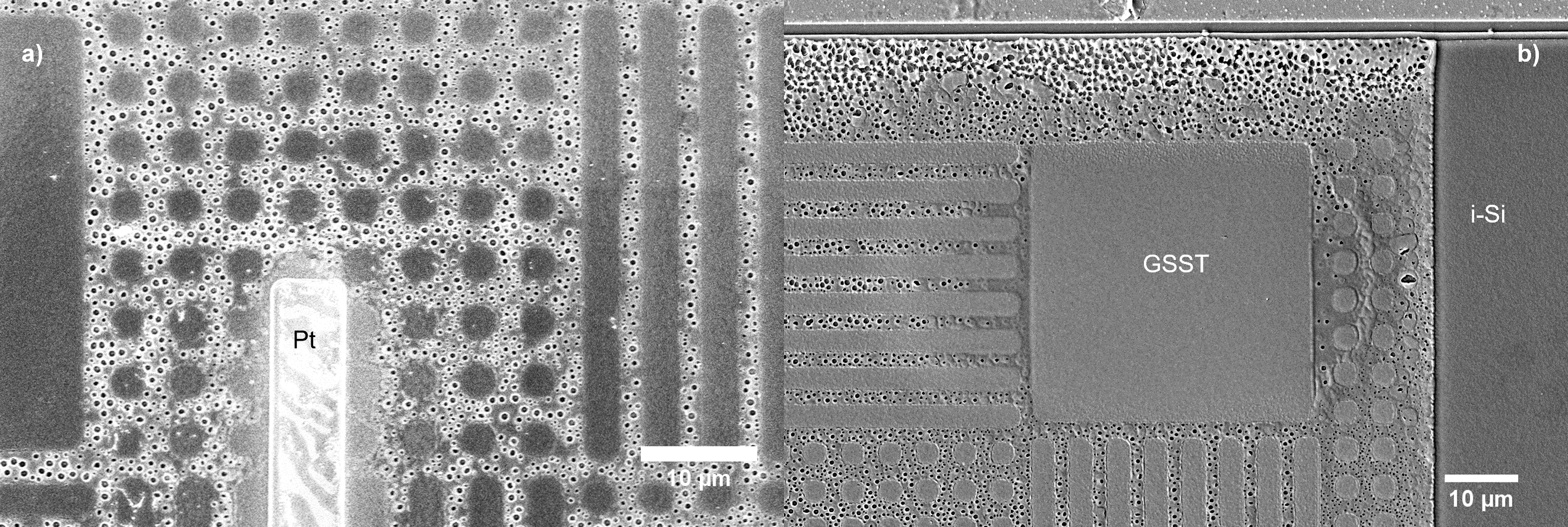}
    \caption{SEM micrographs of a device encapsulated with PECVD SiN$_x$ after cycling at (a) the center and (b) the edge of the device, highlighting the difference in morphology between center to the of the device. The Pt strip in the first panel was protecting a region for TEM.}  
    \label{figSI:HydrogenMorphologyDevice}
\end{figure}

\textbf{Dynamics of PECVD SiN$_x$ film upon cycling}  \newline
A tested device with PECVD silicon nitride displayed the degradation in optical quality of the protective layer (due to the reasons mentioned in the main manuscript) but also showed the recovery of the transparency of the SiN$_x$ as it was cycled (Figure SI \ref{figSI:Opt_Hydro_Evol}). This behavior has not been observed consistently and it is expected to present uncertainty in potential applications but may prove of interest for further development of such PECVD films in terms of observations of kinetics of the system. The size and shape of the delamination/pinhole structures was noticed to evolve both in time (Figure SI \ref{figSI:Opt_Hydro_Evol}) and space (Figure SI \ref{figSI:HydrogenMorphologyDevice}), with a dependence of the size based on the position on the heater. Two potential reasons for that are the inherent thermal profile of the heater and the potential for agglomeration of these voids at edge, leading to fusing of pinholes due to their inability to continue migrating.

\begin{figure}[h!]
    \centering
    \includegraphics[width = 0.5\linewidth]{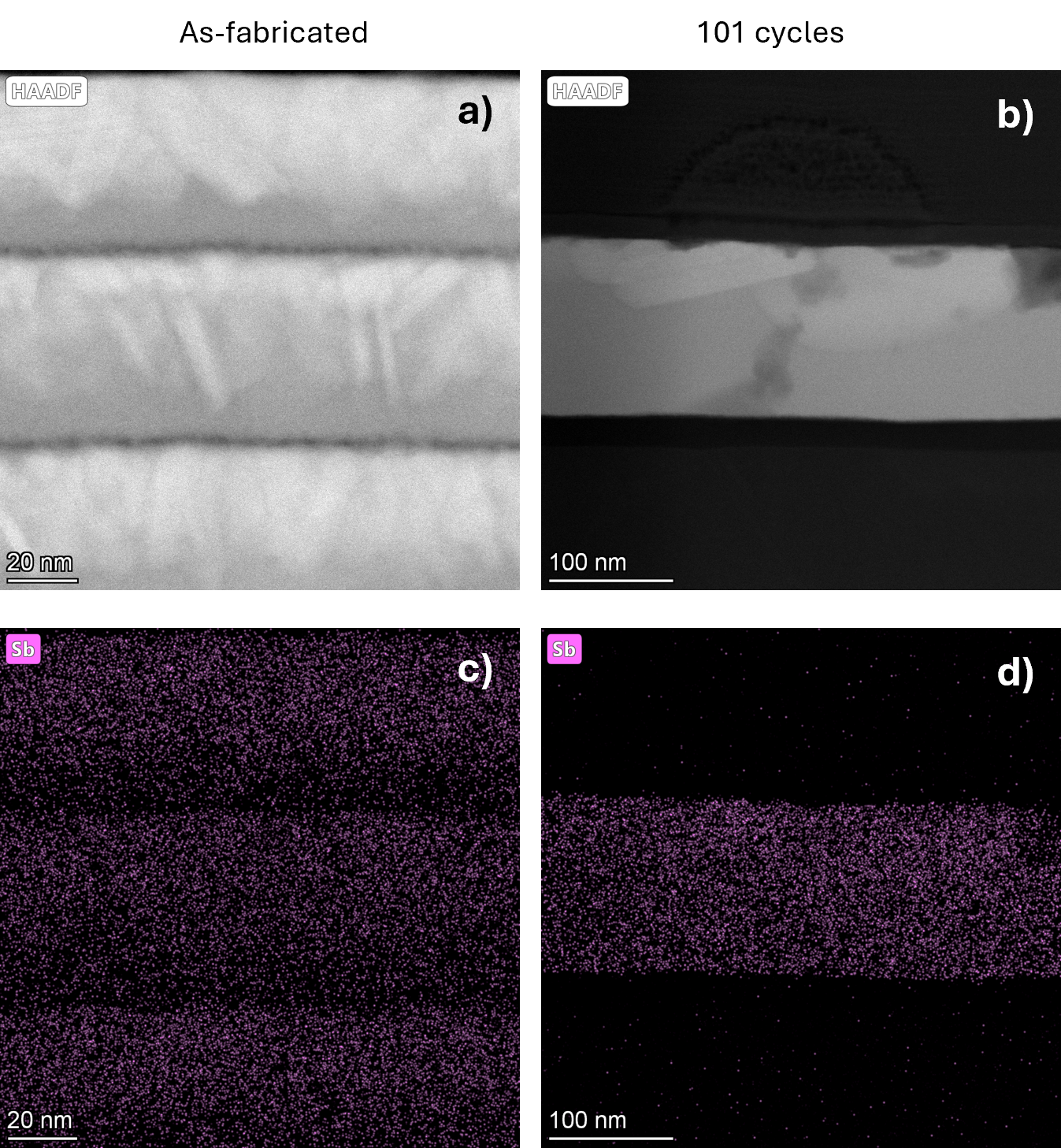}
    \caption{ (a,b) STEM images and corresponding (c,d) EDS mapping of Sb of a control sample (a,c) which was deposited with 150 nm of GSST via 3 depositions (3 $\times$ 50 nm) vs (b,d) another sample from the same chip that was cycled for 101 cycles showing the erasure of the initial triple layer profile. }  
    \label{figSI:Sb_ThreeLayers_TEM}
\end{figure}

\textbf{Erasure of initial PCM state} \newline
A device where three sequential depositions of GSST were performed was cycled and it can be seen that relatively early in its lifetime (100 cycles), the initial morphology and composition gradient is erased by the applied thermal stimuli Figure SI \ref{figSI:Sb_ThreeLayers_TEM}, giving information about the behavior of the material earlier in its life-cycle. Sb was highlighted in the figures for clarity.

\begin{figure}[h!]
    \centering
    \includegraphics[width = 0.5\linewidth]{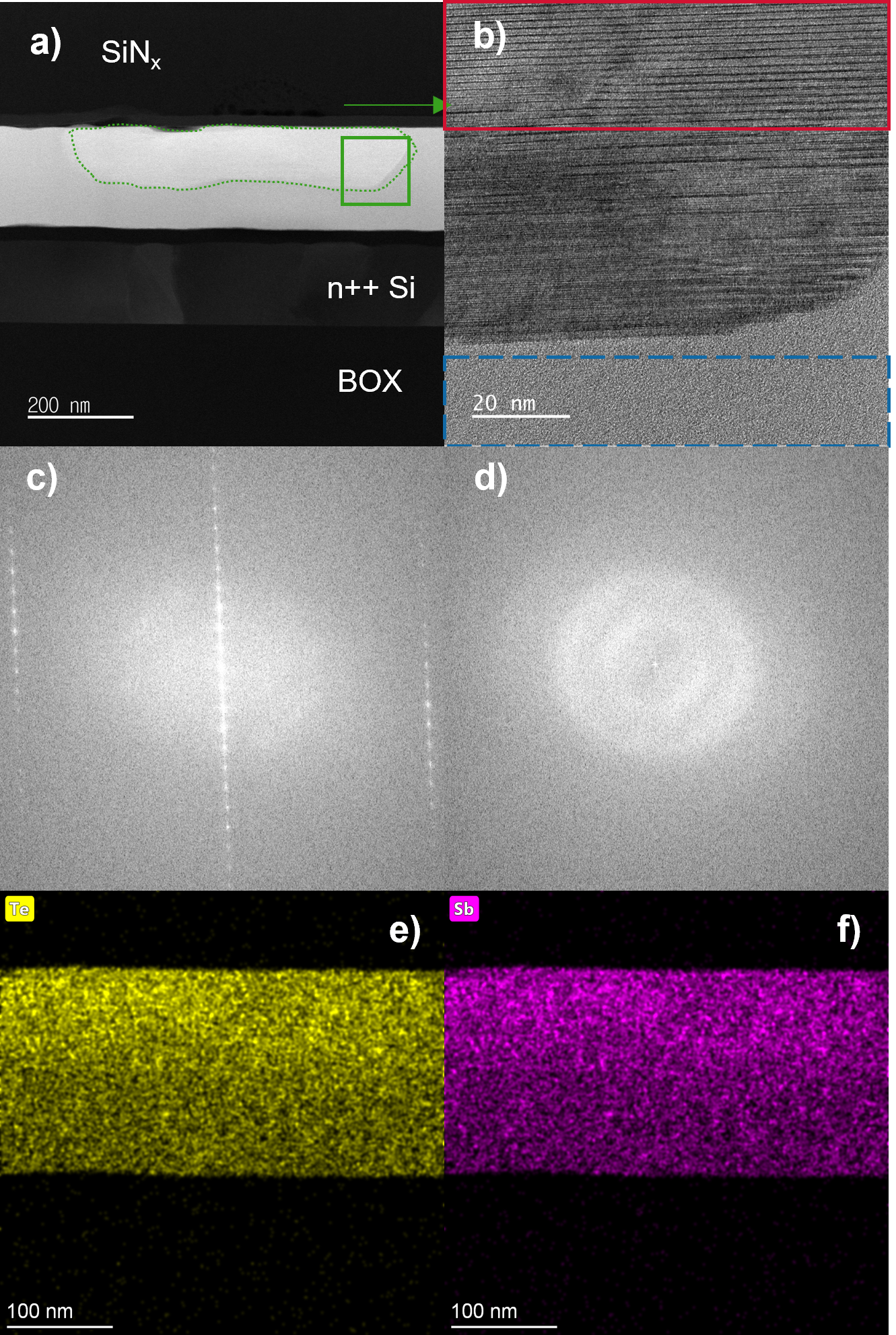}
    \caption{ (a) STEM image of a different region of the lamellae shown in the main manuscript in Figure \ref{fig5:TEM} (i-l), highlighting one of the few crystalline islands remaining after the homogenization pulse with a dashed line. (b) HRTEM micrograph of the square region  in (a) highlighting the lamellar structure of the crystalline island. The solid red rectangle and dashed blue rectangle show approximate regions of analysis for Fourier transformation for (c) and (d) respectively. (c) shows a crystalline patter for the island while the region below it, with its transform in (d) shows an amorphous structure. This region is highlighted as an outlier of this sample and the presence of the crystalline region is associated with its location at the edge of the heater, near a heat sink. The EDS maps of Te and Sb for the island region are shown in (e) and (f) respectively. }  
    \label{figSI:Island_After_Homogenization}
\end{figure}

\textbf{PCM behavior post-homogenization} \newline
A region in the homogenized GSST structure from the main manuscript (Figure \ref{fig5:TEM} i-l) is highlighted for the presence of a crystalline island. Below the crystalline island, the Fourier transform of the HRTEM image shows the typical amorphous rings but  the region at the top shows (in Fourier transform) clearly well separated peaks, congruent with a crystalline region. This island is heavier in Sb and Te (Figure SI \ref{figSI:Island_After_Homogenization}), with a composition measured around Ge 23.0\%, Sb 23.4\%, Se 47.2\% and Te 6.4\% in the homogenized region vs  Ge 12.2\%, Sb 33.3\%, Se 42.2\% and Te 12.3\% in the island.

\begin{figure}[h!]
    \centering
    \includegraphics[width = 0.8\linewidth]{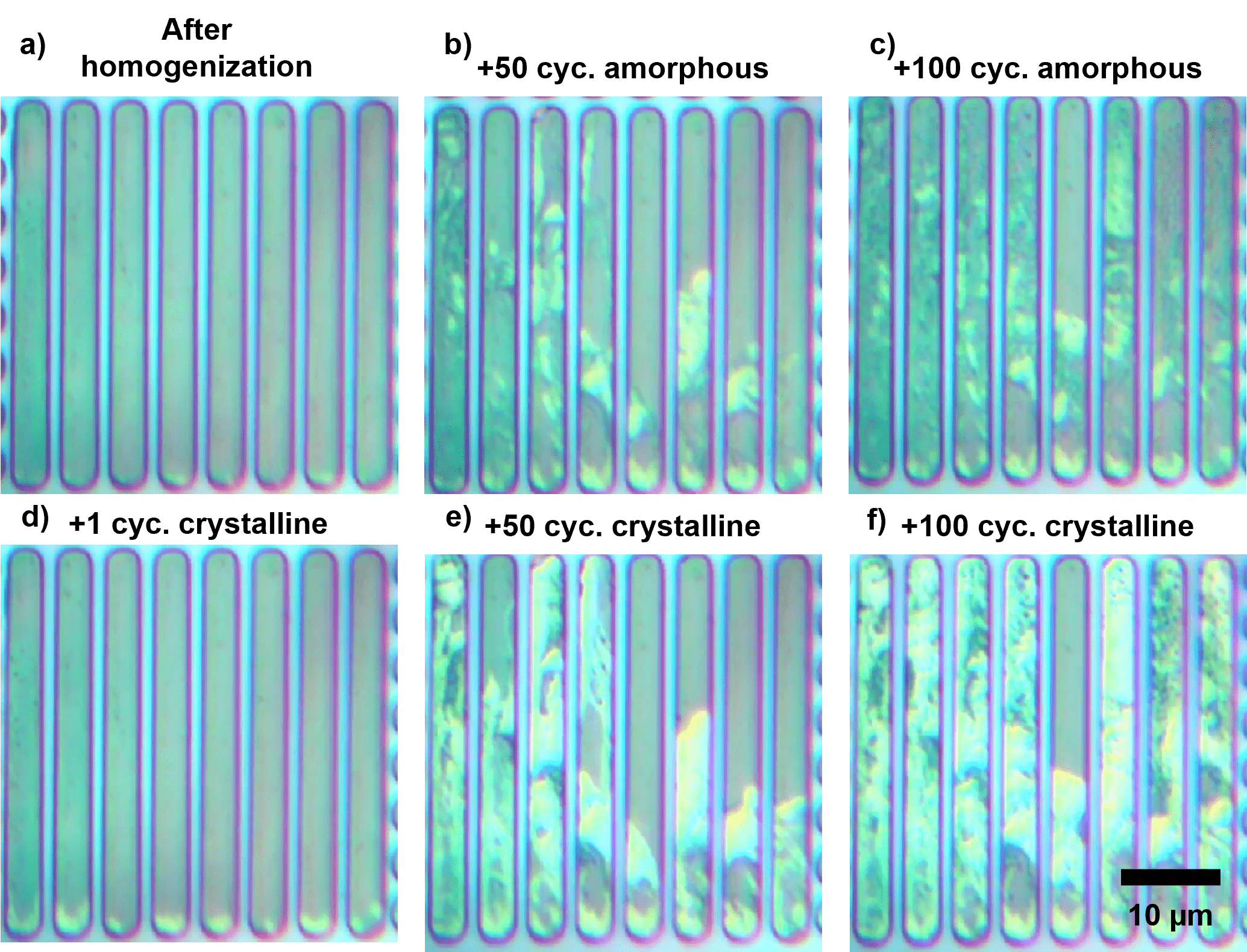}
    \caption{A device after 18 000 cycles that was subjected to a homogenization pulse shown in its nominally amorphous (a, b, c) and crystalline state (d, e, f), immediately after the application of the pulse (a),  after crystallization in its first cycle (d), after 50 cycles (b, e) and 100 cycles (c, f). The weaker applied amophization pulses are sufficient to trigger a large contrast change, similar to original optical contrast, but they are not sufficient to fully erase the crystalline structure. As a result, upon every cycle, the semi-crystalline front propagates from the bottom, cold edge of the gratings, similar to columnar growth in classical systems. A level of heterogeneous morphology is typically present after homogenization (e,f) which may be linked to re-initiation of phase segregation.  }  
    \label{figSI:Homogenization_Optical}
\end{figure}

The homogenization process can be done repeatedly in order to fully erase the prior crystalline structure and drive the PCM back to a homogeneous state. A device that was tested as such is shown in SI Figure \ref{figSI:Homogenization_Optical}. After the initial homogenization, it displays a dark slightly yellow-green but as repeated typical pulses are applied to cycle it, the switching area increases as a crystalline or semi-crystalline structure is partially maintained even during the nominal amorphous form (showing a slightly bluish tone). The crystalline state shows as highly reflective after homogenization. A large contrast can be seen by comparing SI Figure \ref{figSI:Homogenization_Optical} (c) vs (f) but it is expected that the contrast window will decrease.

\begin{figure}
\textbf{Table of Contents}\\
\medskip
  \includegraphics{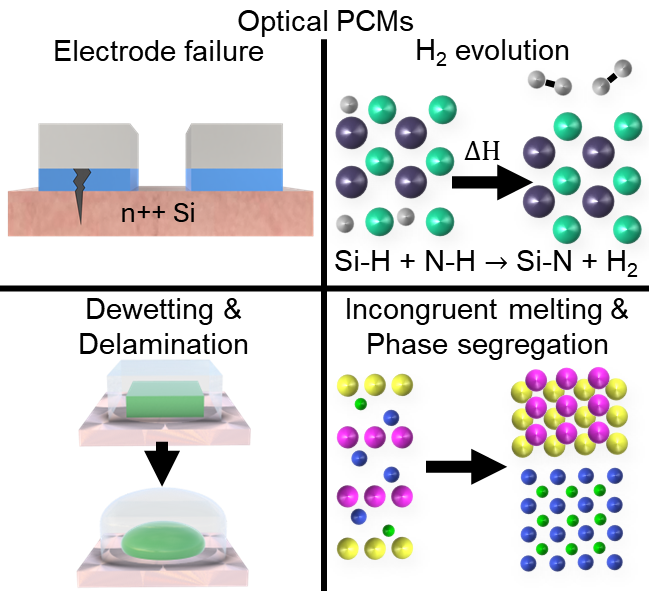}
  \medskip
  \caption*{Chalcogenide PCMs have gathered major interest as a key enabling material for reconfigurable optics and photonics. Using PCMs in optical devices require new considerations distinct from those for electronic phase change memories that need to be addressed towards improved reliability. We identify failure mechanisms pertaining to optical PCM devices and implement methods to significantly boost device lifetime.}
\end{figure}

\end{document}